\documentclass[twocolumn,preprintnumbers,amsmath,amssymb]{revtex4}
\usepackage{graphicx} % include figure files
\usepackage{setspace} % include figure files
\usepackage{dcolumn} % align table columns on decimal point
\usepackage{bm} % bold math
\usepackage{natbib}
\usepackage{hyperref}
\hypersetup{colorlinks=true,linkcolor=blue,citecolor=blue,urlcolor=blue}
\usepackage{epsfig}

\begin{document}
\title{Thermoelectric properties of high-entropy rare-earth cobaltates}
\author{Ashutosh Kumar} 
\author{Diana Dragoe} 
\author{David Bérardan} 
\author{Nita Dragoe\footnote{Email : nita.dragoe@universite-paris-saclay.fr}} 

\affiliation{ICMMO (UMR CNRS 8182), University of Paris-Saclay, Orsay 91405, France}

\begin{abstract}
High-entropy concept introduced with a promising paradigm to obtain exotic physical properties has motivated us to explore the thermoelectric properties of Sr-substituted high-entropy rare-earth cobaltates i.e., (LaNdPrSmEu)$_{1-x}$Sr$_x$CoO$_3$ (0 $\leq$ x $\leq$ 0.10). The structural analysis of the samples synthesized using the standard solid-state route, confirms the orthorhombic structure with the \textit{Pbnm} space group. The Seebeck coefficient and electrical resistivity decrease with rising Sr concentration as well as with an increase in temperature. The multiple A-site ions in high-entropy rare-earth cobaltates result in an improved Seebeck coefficient ($\alpha$) compared to La$_{0.95}$Sr$_{0.05}$CoO$_3$, associated with a decrease in the Co-O-Co bond angle, which further enhances the power factor. The random distribution of cations at the rare-earth site results in a significant lowering of phonon thermal conductivity. As a result, a maximum figure of merit (zT) of 0.23 is obtained at 350\,K for (LaNdPrSmEu)$_{0.95}$Sr$_{0.05}$CoO$_3$, which is one of the highest values of zT reported at this temperature for oxide materials. This study shows promise to decouple thermoelectric parameters using the high-entropy concept in several materials.
\end{abstract}
\maketitle
High-entropy oxides (HEOx) are an emerging class of compounds, promising for several exotic electronic, magnetic, and thermal properties.\cite{R1, R2, R3, R4, R5} These materials are now thoroughly studied and several interesting properties have already been reported, including for example long-range antiferromagnetic order, cathode materials for sodium-ion batteries, colossal dielectric constant, superionic conductivity, large lithium storage, and proton conductors.\cite{R6,R7, R8, R9, R10, R11, R12} One of the interests in these materials is related to the statistical distribution of several cations on one single crystallographic site, which should increase phonon scattering and hence lower the lattice thermal conductivity ($\kappa_{ph}$). Several reports on the influence of high-entropy on thermoelectric (TE) properties of n-type materials appeared recently. Lou et al., showed a low thermal conductivity ($\kappa$=1.89 W/m-K @ 873K) in Sr$_{0.9}$La$_{0.1}$(Zr$_{0.25}$Sn$_{0.25}$Ti$_{0.25}$Hf$_{0.25}$)O$_3$, whereas an ultralow $\kappa$=0.7 W/m-K @ 1100 K is shown for high-entropy Sr(Ti$_{0.2}$Fe$_{0.2}$Mo$_{0.2}$Nb$_{0.2}$Cr$_{0.2}$)O$_3$ oxide.\cite{R13, R14, R15, R16}\\
In general, the potential of a TE material requires a high Seebeck coefficient ($\alpha$), low electrical resistivity ($\rho$), and low thermal conductivity ($\kappa$) to realize a high figure of merit (zT), which depends on materials parameters as zT=($\alpha^2$T/($\rho.\kappa$), where T is the absolute temperature and $\kappa$ consists of electronic ($\kappa_e$) and lattice ($\kappa_{ph}$) thermal conductivity i.e., $\kappa$ = $\kappa_e$ + $\kappa_{ph}$. The strong correlation between these TE parameters hinders the reach of high zT in a material. High-entropy oxides may decouple electron and phonon transport in a material as (i) multi-cation substitution at one site having large lattice distortion and mass fluctuation may enable the strong reduction of $\kappa_{ph}$,\cite{R15, R18}(ii) high-entropy compounds tends to form a high-symmetry phase that is beneficial for band degeneracy and hence may enhance $\alpha$,\cite{R19} (iii) heavy-doping may be possible in the high-entropy phase due to possibly larger solubility limits that may optimizing $\rho$ in a system.\\ 
Rare-earth (RE) cobaltates, a prototypical perovskite oxide (ABO$_3$, where A is a rare-earth element, B is transition metal and O is oxygen), are strongly correlated electron systems having different charge states along with different spin states of cobalt.\cite{R20, R21, R22, R23, R24} The degenerate electronic state of Co$^{3+}$ (Co$^{4+}$) is a result of the correlation between Hund’s rule coupling and crystal field in high spin (HS), intermediate spin (IS), and low spin (LS) states. In the past, the electronic properties in the rare-earth cobaltates have been modified through single, and/or co-substitution at the RE and at Co sites in RECoO$_3$.\cite{R20, R25, R26, R27} Wang \textit{et al.,} demonstrated the correlation between the structural distortion through Sr and Ca substitution at the La site on the TE properties (zT of 0.10 at 300 K) in LaCoO$_3$.\cite{R27} Further, Sr and Mn co-substitution in LaCoO$_3$ resulted in a zT of 0.14 at 480 K via tuning the spin-state transition temperature.\cite{R26} The $\alpha$ in LaCoO$_3$ has a strong dependence on the spin-state of cobalt and temperature. The spin-states of cobalt have thermally activated transition near 100 K from LS to IS and near 500 K from IS to HS, showing metallic nature at higher temperatures ($>$600K) and resulting in a constant Seebeck coefficient $\sim$37 $\mu$V/K irrespective of the nature and level of substitution \cite{R28} and hence becomes less promising for TE properties at higher temperatures. Further, $\kappa_{ph}$ in rare-earth cobaltates has a dominating role in $\kappa$ that reduces zT in these oxide materials at lower temperatures.\cite{R29, R30} The reduction in $\kappa_{ph}$ in LaCoO$_3$ can be obtained via nanostructurization (scatter mid and long-wavelength phonons), doping (scatter short-wavelength phonons), and composite approach. However, reduction in $\kappa_{ph}$ also results in reduced electrical conductivity ($\sigma$) due to scattering of charge carriers, if the length scale is not optimized properly. The intertwined dependence of grain size on electronic and thermal conductivity limits the desired TE parameters. The HEO$_x$ concept, hence, can be used to induce multi-phonon scattering in oxide materials to reduce the dominating $\kappa_{ph}$ and hence may be beneficial for improving TE properties. Herein, we aim to explore the influence of multi-cation distribution at the rare-earth site on the thermoelectric properties of p-type rare-earth cobaltates.
\\
\\
High-entropy rare-earth cobaltates (LaNdPrSmEu)$_{1-x}$Sr$_x$CoO$_3$ (x=0.00, 0.05, and 0.10) were synthesized using standard solid-state reaction methods. The stoichiometric amount of La$_2$O$_3$ (Alfa Aesar,99.9\%) Nd$_2$O$_3$ (Alfa Aesar, 99.9\%) Pr$_2$O$_3$ (Alfa Aesar, 99.9\%), Sm$_2$O$_3$ (Alfa Aesar, 99.9\%), Eu$_2$O$_3$ (Alfa Aesar, 99.9\%), SrCO$_3$ (Alfa Aesar, 99.9\%) and Co$_3$O$_4$ (Alfa Aesar, 99.99\%) were mixed using a planetary ball milling (Fritsch Pulverisette 7 Premium Line) for 12 cycles (350 rpm, 5 min on and 1 min off time). The rare-earth precursors were heated overnight at 1173\,K before synthesis to avoid any water adsorption in the powders. The mixed precursors were heated at 1573\,K for 20 hours with a heating and cooling rate of 200 K/hour. The calcined powders were mixed in an agate mortar pestle to make a homogeneous mixture. Further, the mixed powders were uniaxially consolidated into pellets and sintered at 1623 K for 20 hours with a 200K/hour heating and cooling rate. A similar synthesis condition was followed to prepare the La$_{0.95}$Sr$_{0.05}$CoO$_3$ sample. The structural characterization of the samples was performed using X-ray diffraction (using a Panalytical X'Pert diffractometer with a Ge(111) incident monochromator, a copper tube (K$_{\alpha1}$ radiation), and a fast detector (X'celerator)). Surface morphology and chemical compositions were confirmed using scanning electron microscopy (SEM-FEG Zeiss Sigma HD) equipped with energy-dispersive X-ray spectroscopy (EDS). Further, X-ray photoelectron spectroscopy (XPS) measurements were performed on a Thermo Scientific instrument with a monochromatic Al-K$_{\alpha}$ X-ray source (energy 1486.68 eV) and a hemispherical analyzer. Powder samples were used for the measurement. The base pressure was around 5$\times$10$^{-9}$ mbar and the diameter of the X-ray beam spot was 400 $\mu$m, corresponding to an irradiated surface of approximately 1 mm$^2$. The hemispherical analyzer was operated at 0$^\circ$ take-off angle in the Constant Analyzer Energy (CAE) mode. Wide scan spectra were recorded at pass energy of 200 eV and an energy step of 1 eV while narrow scan spectra were recorded at a pass energy of 50 eV and an energy step of 0.1 eV. Charge compensation was achieved by means of a “dual beam” flood gun, using low-energy electrons (5 eV) and argon ions. The binding energy scale was calibrated on the neutral carbon set at 285 eV.  The electrical resistivity ($\rho$) and Seebeck coefficient ($\alpha$) were measured using homemade instruments over a wide temperature range of 300\,K-1000\,K in a standard four-probe configuration.\cite{R31} The transport measurements were performed in both heating and cooling modes and the results are identical. However, for clarity of presentation, only measurement data during heating are presented. The thermal conductivity of the sample was calculated using the following relation: $\kappa$=D$\rho_s$C$_p$. The thermal diffusivity (D) was measured using the laser flash technique (Netzsch LFA 427), and sample density ($\rho_s$) is calculated using sample mass and its geometric volume. The specific heat capacity (C$_p$) was used from the literature to calculate $\kappa$.\cite{R32}\\ 
\\
The powder X-ray diffraction patterns (PXRD) indicate the single-phase formation of high-entropy (LaNdPrSmEu)$_{1-x}$Sr$_x$CoO$_3$ samples (Fig.~S1(a)). It is worth noting that the crystal structures for individual rare-earth cobaltates (RECoO$_3$) are different (i.e., RE=Nd, Pr, Sm possesses orthorhombic structure, RE=La crystallizes in rhombohedral structure and RE=Eu forms a cubic phase)\cite{R30}; however, the high-entropy samples stabilize in orthorhombic structure with \textit{Pbnm} space group. Further, the nature of the diffraction pattern remains the same with the Sr substitution at the RE site in high-entropy samples. The PXRD is further analyzed using the Reitveld refinement\cite{R25aa} via Fullprof software and the refinement pattern is shown in Fig.~S1(b). The decrease in lattice parameters (Table SI) obtained in the Sr substituted high-entropy (LaNdPrSmEu)$_{1-x}$Sr$_x$CoO$_3$ samples is consistent with the difference in ionic radii of Sr and rare-earth elements. Further, the diffraction pattern of La$_{0.95}$Sr$_{0.05}$CoO$_3$ shows the rhombohedral structure with the \textit{R-3c} space group, and the corresponding refinement pattern is shown in Fig.~S2. The PXRD for simple perovskite (La$_{0.95}$Sr$_{0.05}$CoO$_3$) and high-entropy (LaNdPrSmEu)$_{0.95}$Sr$_{0.05}$CoO$_3$ perovskite is shown in Fig.~1(a). The corresponding crystal structure, prepared from the structural information obtained from Rietveld refinement, using Vesta software \cite{R25aaaa} is shown in Fig.~1(b-c). The La$_{0.95}$Sr$_{0.05}$CoO$_3$ crystallizes in a rhombohedral structure with two formula units. The structural motif of this compound is a nearly perfect CoO$_6$ octahedron. The rhombohedral distortion from perfect cubic perovskite depicts the deformation along the body diagonal to decrease the Co-O-Co bond angle below 180o.  In (LaNdPrSmEu)$_{0.95}$Sr$_{0.05}$CoO$_3$, the main crystallographic motif CoO$_6$ octahedron is rotated about the ab plane and titled with respect to the c-axis in the orthorhombic configuration. The deviation from the cubic symmetry may be theoretically understood by calculating Goldschmidt’s tolerance factor (t) for the prepared systems. Since the rare-earth site is populated with the five-different rare-earth elements with smaller ionic radii than La, the tolerance factor further deviates in HEO$_x$ (t=0.93 for La$_{0.95}$Sr$_{0.05}$CoO$_3$ and t=0.92 for (LaNdPrSmEu)$_{0.95}$Sr$_{0.05}$CoO$_3$) from an ideal cubic structure (t=1).\\ 
The compact surface morphology is seen in the scanning electron microscopy (SEM) images (Fig.~S3), which results in a high relative density (~86-89\%) for all the samples and is attributed to the high sintering temperature.  Further, the atomic percentage of each element obtained from the energy-dispersive X-ray spectroscopy (EDS) measurement is in agreement with the stoichiometric amount of the synthesized samples (Fig.~S3). The homogenous distribution is also observed in the elemental mapping of these samples (Fig.~S3).
\\
\begin{figure}
\centering
\includegraphics[width=0.99\linewidth]{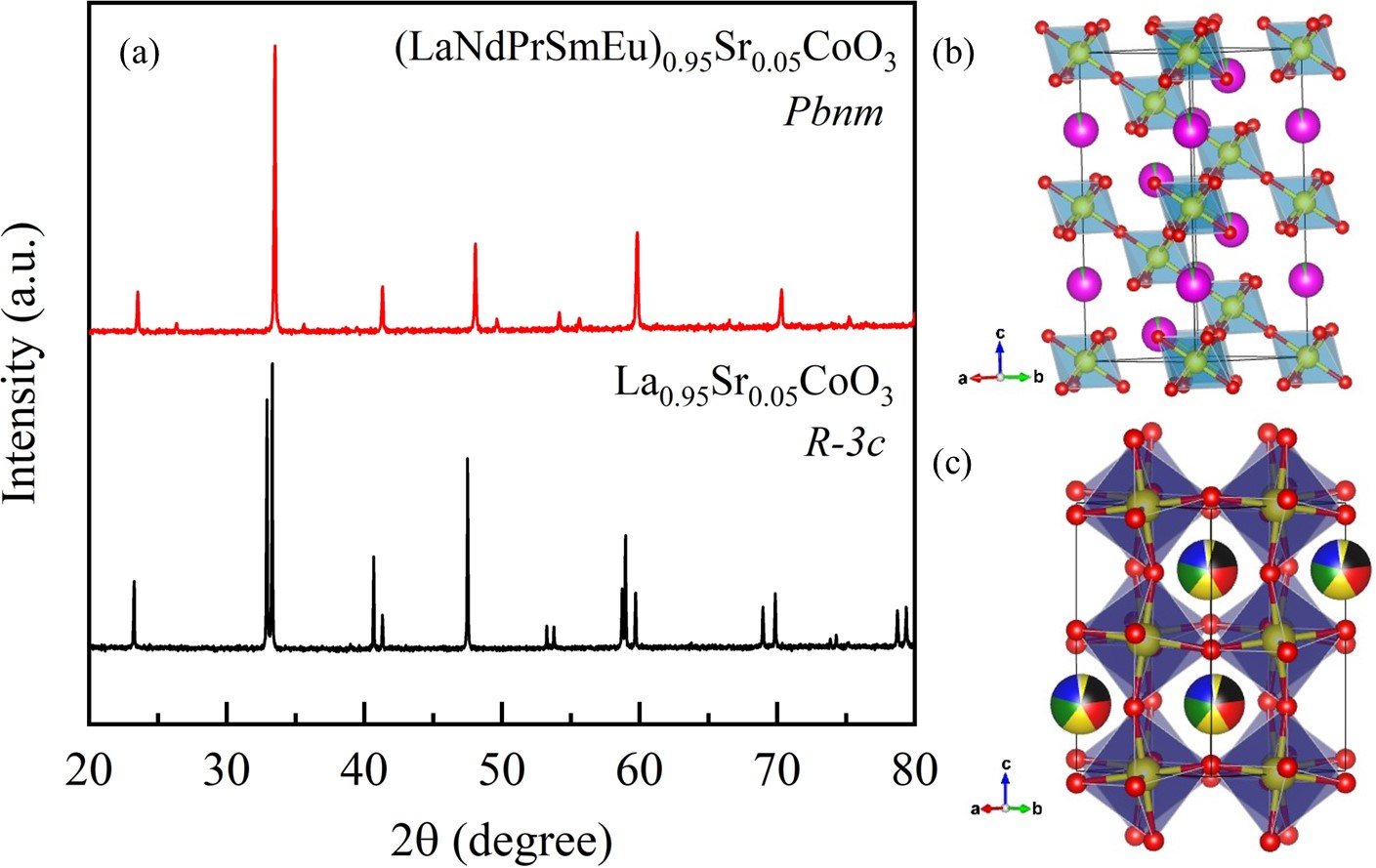}
\caption{(a) XRD pattern of La$_{0.95}$Sr$_{0.05}$CoO$_3$ and (LaNdPrSmEu)$_{0.95}$Sr$_{0.05}$CoO$_3$ is shown. The crystal structure using Vesta software is shown for (b)  La$_{0.95}$Sr$_{0.05}$CoO$_3$, and (c) (LaNdPrSmEu)$_{0.95}$Sr$_{0.05}$CoO$_3$. The structural parameters are obtained from the Rietveld refinement of the XRD pattern.}
\end{figure}
The valence states of the elements present in the high-entropy rare-earth cobaltates have been further investigated using X-ray photoelectron spectroscopy (XPS) measurement, shown in Fig.~S4. Fig.~S4(a) shows the presence of all the expected elements in the full scan range for (LaNdPrSmEu)$_{1-x}$Sr$_x$CoO$_3$, consistent with the SEM-EDS measurements. The rare-earth elements gave identical core-level spectra for all the samples. The oxidation state for the rare-earth remains the same for pristine and Sr-substituted samples (see Fig.~S5). The Sr3d core-level spectrum is shown in Fig.~S4(b), it depicts the increases in the intensity corresponding to Sr-3d with Sr concentration in the system. The core-level spectrum of Co2p consists of two main peaks which are the spin-orbit coupling components  Co2p$_{3/2}$ at 780.6 $\pm$ 0.2 eV and Co2p$_{1/2}$ at 795.8 $\pm$ 0.2 eV amd two small satellite peaks. (Fig.~S4(c)). The seperation between Co2p3/2 and its satellite is about 9.6 eV, indicating the presence of the Co3+ state. \cite{R26aa} The area corresponding to Co2p$_{3/2}$ and Co2p$_{1/2}$ increases with Sr concentration, indicating an increased contribution of Co$^{4+}$.\cite{R27aa} The O1s core-level spectrum is shown in Fig.~S4(d). The peaks corresponding to 529.4$\pm$0.2 eV and 531.1$\pm$0.2 eV can be ascribed to lattice oxygen (O$_L$) and surface-absorbed oxygen (O$_A$), respectively. The increase in O$_A$ intensity indicates that the holes originating due to an increase in Sr concentration in the system may enhance the surface oxygen absorption to maintain charge neutrality.\\ 
The Seebeck coefficient ($\alpha$) as a function of temperature is shown in Fig.~2(a). A positive value of $\alpha$ is obtained for all the samples, depicting dominating p-type conduction in the system. It is worth noting that the sample with x=0.00 also depicts a positive and large value of $\alpha$, unlike LaCoO$_3$, and NdCoO$_3$.\cite{R33, R34, R35} The large value of $\alpha$ in the LaCoO$_3$ system is attributed to their strong electronic correlation and large configurational entropy that arises from several spin-states and orbital degeneracy of cobaltates.\cite{R20} The $\alpha$ at 300\,K for (LaNdPrSmEu)$_{1-x}$Sr$_x$CoO$_3$ is $\sim$450 $\mu$V/K for x=0.00 and it decreases with an increase in temperature. The $\alpha$ for different rare-earth cobaltates is different and is attributed to their ionic sizes that tune the Co-O-Co bond angle and affects the overlap between the O p orbital and the Co3d orbital and hence modify the electronic bandwidth.\cite{R37} It should be noted that the $\alpha$ for (LaNdPrSmEu)$_{0.95}$Sr$_{0.05}$CoO$_3$ is larger than that of La$_{0.95}$Sr$_{0.05}$CoO$_3$ ($\sim$275 $\mu$V/K at 300K). Indeed the $\alpha$ for other rare-earths (like Nd, Pr, Sm, etc) cobaltates is higher than that of LaCoO$_3$.\cite{R36} This change in $\alpha$ for different rare-earth cobaltates is attributed to the variation in the Co-O-Co bond angle which changes with the ionic radii of RE elements. \cite{R37} Since the high-entropy sample (LaNdPrSmEu)$_{0.95}$Sr$_{0.05}$CoO$_3$ consists of multiple rare-earth elements having small ionic radii than La, the effective bond angle is lower which results in a higher Seebeck coefficient as compared to La$_{0.95}$Sr$_{0.05}$CoO$_3$. The Sr-substitution at the rare-earth site reduces $\alpha$ (from 300 $\mu$V/K for x=0.05 to 155 $\mu$V/K for x=0.10), similarly to LaCoO$_3$, indicating the increase of carrier concentration, as also evident from the XPS measurement.\\

\begin{figure}
\centering
\includegraphics[width=1.0\linewidth]{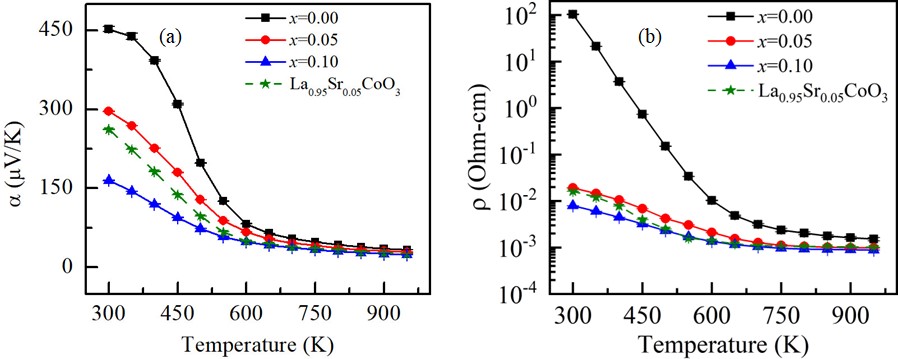}
\caption{Figure 2: Seebeck coefficient ($\alpha$), electrical resistivity ($\rho$) as a function of temperature for (LaNdPrSmEu)$_{1-x}$Sr$_x$CoO$_3$.}
\end{figure}
Temperature-dependent electrical resistivity ($\rho$) of (LaNdPrSmEu)$_{1-x}$Sr$_x$CoO$_3$ is shown in Fig.~2(b). The $\rho$ decreases with an increase in the temperature (d$\rho$/dT$<$0) for all the samples, indicating semiconducting behavior. The $\rho$ decreases with Sr substitution for all the samples. It is ascribed to the generation of charge carriers due to Sr$^{2+}$ substitution at the RE site.\cite{R36} It is seen from Fig.~2(b) that $\rho$ of high-entropy (LaNdPrSmEu)$_{0.95}$Sr$_{0.05}$CoO$_3$ is higher than that of the La$_{0.95}$Sr$_{0.05}$CoO$_3$. This is understood as follows:  the substitution of the RE site with smaller ions (from La to Eu) decreases the effective ionic radii reducing the Co-O-Co bond angle and hence decreases the overlap between O p-orbital and Co 3p orbitals which results in higher $\rho$. It is also noted that in literature, $\rho$ for different rare-earth cobaltates changes by a few orders of magnitude\cite{R30}; however, in the present study, $\rho$ for the high-entropy (LaNdPrSmEu)$_{0.95}$Sr$_{0.05}$CoO$_3$ for x=0.05 is slightly higher than that of La$_{0.95}$Sr$_{0.05}$CoO$_3$, and this may be due to the random distribution of these RE elements at the A site. The increase in $\rho$ for the high-entropy (LaNdPrSmEu)$_{0.95}$Sr$_{0.05}$CoO$_3$ system is consistent with the rise in $\alpha$, which results in an improved power factor ($\alpha^2$/$\rho$) for the high-entropy system (shown in Fig.~S6). Thus, it indicates that the high-entropy configuration in rare-earth cobaltates results in a higher $\alpha$ and hence improves $\alpha^2$/$\rho$.\\
\begin{figure}
\centering
\includegraphics[width=1.0\linewidth]{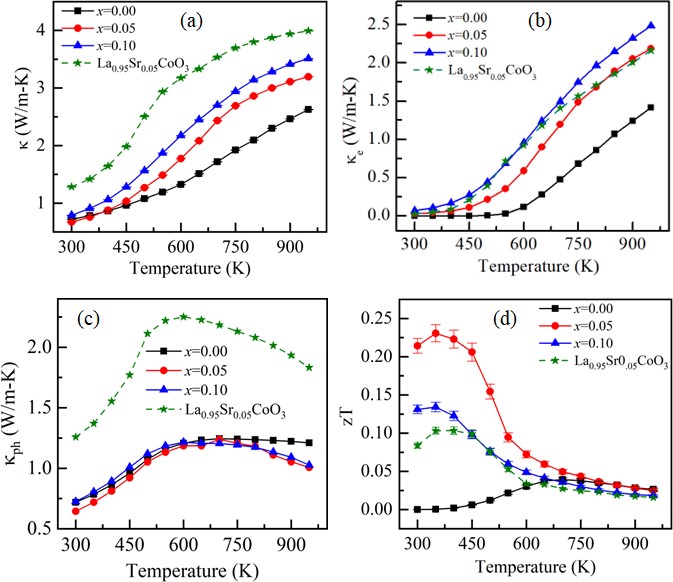}
\caption{(a) Total thermal conductivity ($\kappa$), (b) electronic thermal conductivity ($\kappa_e$), (c) phonon thermal conductivity ($\kappa_{ph}$), and (d) figure of merit (zT) as a function of temperature for (LaNdPrSmEu)$_{1-x}$Sr$_x$CoO$_3$. A comparison of these parameters for simple perovskite La$_{0.95}$Sr$_{0.05}$CoO$_3$ is also shown.}
\end{figure}
\\
Total thermal conductivity ($\kappa$) as a function of temperature is shown in Fig.~3(a). $\kappa$ increases with an increase in temperature for all the samples. This increase may be attributed to the rise in $\sigma$ in the system. It is noted that the $\kappa$ value for the high-entropy sample is significantly lower than that of the simple perovskite sample. As discussed earlier, $\kappa$, in general, results from two contributions: electronic thermal conductivity ($\kappa_e$) and phonon thermal conductivity ($\kappa_{ph}$). In order to depict the change in $\kappa$ in two different configurations i.e., in high-entropy samples and simple perovskite sample, the $\kappa_e$ has been first estimated using Wiedemann Franz law: $\kappa_e$ =L$\sigma$T, where Lorenz number ($L$) is calculated using the equation: L=1.5+exp[-$\alpha$/116].\cite{R38} $\kappa_e$ and $\kappa_{ph}$ as a function of temperature for all the samples are shown in Fig.~3(b) and Fig.~3(c), respectively. The rise in $\kappa_e$ agrees with the increase in $\sigma$. The $\kappa_{ph}$ decreases at higher temperatures, attributed to the decrease in the mean free path of phonons at higher temperatures, for all the samples. The novel observation of this study lies in the significantly reduced $\kappa_{ph}$ for the high entropy samples as compared to the “simple” perovskite (La$_{0.95}$Sr$_{0.05}$CoO$_3$), consistent with the fact that the random distribution of different rare-earth elements with different ionic radii increases phonon scattering and hence reduce the $\kappa_{ph}$. The $\kappa_{ph}$  at 300\,K for La$_{0.95}$Sr$_{0.05}$CoO$_3$ is 1.28 W/m-K which is reduced to 0.67 W/m-K in high-entropy (LaNdPrSmEu)$_{0.95}$Sr$_{0.05}$CoO$_3$.\\
The figure of merit (zT) calculated using the measured thermoelectric parameters is shown in Fig. 3(d). The zT for the pristine high-entropy (LaNdPrSmEu)CoO$_3$ sample is poor at 300 K due to its high electrical resistivity. However, with Sr substitution, the zT improves at lower temperatures and reaches a maximum value of 0.23 at 350K in (LaNdPrSmEu)$_{0.95}$Sr$_{0.05}$CoO$_3$, which is significantly larger than the simple perovskite sample (La$_{0.95}$Sr$_{0.05}$CoO$_3$) at the same temperature. The zT value obtained for La$_{0.95}$Sr$_{0.05}$CoO$_3$ in the present study is consistent with the previous reports.\cite{R26, R28} This rise in zT for the high-entropy sample can be attributed to the increase in the Seebeck coefficient and the significant reduction in the phonon thermal conductivity in a high-entropy configuration. This is one of the highest values of zT obtained in oxide materials at this temperature range. \cite{R39}\\
\\
In summary, high-entropy rare-earth cobaltates have been synthesized using standard solid-state reactions. The X-ray diffraction followed by Rietveld refinement confirms the single-phase formation. The high-entropy configuration results in orthorhombic structure ((LaNdPrSmEu)$_{1-x}$Sr$_x$CoO$_3$) in contrast to the rhombohedral structure in the simple perovskite (La$_{0.95}$Sr$_{0.05}$CoO$_3$) sample. The change in structural configuration results in a decrease in the Co-O-Co bond angle, which improves the Seebeck coefficient in high-entropy samples. The multi-cation distribution at rare-earth sites results in a significant reduction in phonon thermal conductivity, which is further lowered with Sr addition. Simultaneous effect of improved Seebeck coefficient and reduced thermal conductivity results in the enhanced figure of merit (zT) in the high-entropy rare-earth cobaltates. A maximum zT of 0.23±0.02 at 350K is obtained for (LaNdPrSmEu)$_{0.95}$Sr$_{0.05}$CoO$_3$, which is one of the best zT for p-type oxide polycrystalline materials at this temperature.

\newpage
\end{document}